\documentclass[12pt,a4paper]{article}
\usepackage{amsfonts}
\usepackage{amsmath}
\usepackage{amssymb}
\usepackage{indentfirst}
\usepackage{graphics}
\usepackage{graphicx}
\usepackage{epsfig}

\newcommand\ds{\displaystyle}

\newcommand\be{\begin{equation}}
\newcommand\ee{\end{equation}}

\newcounter{offtheorem}
\newenvironment{mytheorem}[1]%
{\begin{trivlist}
     
     \refstepcounter{offtheorem}
     \item[\hspace{\labelsep}\bf #1 \thesection.\arabic{offtheorem}.]}%
{\end{trivlist}}
\newenvironment{prop}{\begin{mytheorem}{Proposition}}{\end{mytheorem}}

\title{\bf Reconstruction Algorithms for Positron Emission
Tomography and Single Photon Emission Computed Tomography and
their Numerical Implementation}
\author{A.S.\ Fokas, A.\ Iserles and V.\ Marinakis
\\ Department of Applied Mathematics and
\\ Theoretical Physics, University of Cambridge
\\ Cambridge, CB3 0WA, United Kingdom}
\date{}

\begin{document}
\maketitle

\begin{abstract}
The modern imaging techniques of Positron Emission Tomography and
of Single Photon Emission Computed Tomography are not only two of
the most important tools for studying the functional
characteristics of the brain, but they now also play a vital role
in several areas of clinical medicine, including neurology,
oncology and cardiology. The basic mathematical problems
associated with these techniques are the construction of the
inverse of the Radon transform and of the inverse of the so called
attenuated Radon transform respectively. We first show that, by
employing mathematical techniques developed in the theory of
nonlinear integrable equations, it is possible to obtain analytic
formulas for these two inverse transforms. We then present
algorithms for the numerical implementation of these analytic
formulas, based on approximating the given data in terms of
cubic splines. Several numerical tests are presented which suggest
that our algorithms are capable of producing accurate reconstruction
for realistic phantoms such as the well known Shepp--Logan phantom.
\end{abstract}

\section{Introduction}
\label{intro}

Positron emission tomography (PET) and single photon emission
computed tomography (SPECT) are two modern imaging techniques with
a wide range of medical applications. Although these techniques
were originally developed for the study of the {\it functional}
characteristics of the brain, they are now used in many diverse
areas of clinical medicine. For example a recent editorial in the
New England Journal of Medicine \cite{koh} emphasized the
importance of PET in oncologic imaging. Other medical applications
of PET and SPECT are presented in \cite{joni}--\cite{beng}.

The first step in PET is to inject the patient with a dose of a
suitable radiopharmaceutical. For example in brain imaging a
typical such radiopharmaceutical is flurodeoxyglucose (FDG), which
is a normal molecule of glucose attached artificially to an atom
of radioactive fluorine. The cells in the brain which are more
active have a higher metabolism, need more energy, thus will
absorb more FDG. The fluorine atom in the FDG molecule suffers a
radioactive decay, emitting a positron. When a positron collides
with an electron it liberates energy in the form of {\it two}
beams of gamma rays travelling in {\it opposite} direction, which
are picked by the PET scanner. SPECT is similar to PET but the
radiopharmaceuticals decay to emit a {\it single} photon.

In both PET and SPECT the radiating sources are inside the body,
and the aim is to determine the distribution $g(x_1,x_2)$ of the
relevant radiopharmaceutical from measurements made outside the
body of the emitted radiation. If $f(x_1,x_2)$ is the $x$--ray
attenuation coefficient of the body, then it is straightforward to
show \cite{natt} that the intensity $I$ outside the body measured
by a detector which picks up only radiation along the straight
line $L$ is given by
\begin{equation}
\label{lineint}
I = \int_L \mathrm{e}^{- \int_{L(x)}f \mathrm{d}s} g \mathrm{d} \tau
\end{equation}
where $\tau$ is a parameter along $L$, and $L(x)$ denotes the section
of $L$ between the point $(x_1,x_2)$ and the detector. The attenuation
coefficient $f(x_1,x_2)$ is precisely the function measured by the
usual computed tomography. Thus the basic mathematical problem in
SPECT is to determine the function $g(x_1,x_2)$ from the knowledge of
the ``transmission'' function $f(x_1,x_2)$ (determined via computed
tomography) and the ``emission'' function $I$ (known from the measurements).

In PET the situation is simpler. Indeed, since the sources eject
particles {\it pairwise} in {\it opposite} directions and the
radiation in opposite directions is measured {\it simultaneously},
equation (\ref{lineint}) is replaced by
\begin{equation}
\label{inew}
I = \int_L \mathrm{e}^{-\int_{L_+(x)}f \mathrm{d}s - \int_{L_-(x)}f
\mathrm{d}s} g \mathrm{d} \tau,
\end{equation}
where $L_+$, $L_-$ are the two half--lines of $L$ with endpoint
$x$. Since $L_++L_-=L$, equation (\ref{inew}) becomes
\[ I = \mathrm{e}^{-\int_L f \mathrm{d} \tau} \int_L g \mathrm{d} \tau. \]
We recall that the line integral of the function $f(x_1,x_2)$ along $L$
is precisely what is known from the measurements in the usual
computed tomography. Thus since both $I$ and the integral of
$f(x_1,x_2)$ are known (from the measurements of SPECT and of
computed tomography respectively), the basic mathematical problem
of PET is to determine $g(x_1,x_2)$ from the knowledge of its
line integrals. This mathematical problem is identical with
the basic mathematical problem of computed tomography.

\subsubsection*{Notation}

\noindent
(i) A point of a line $L$ making an angle $\theta$ with the
$x_1$--axis is specified by the three real numbers
$(\tau,\rho,\theta)$, where $\tau$ is a parameter along $L$,
$-\infty < \tau < \infty$, $\rho$ is the distance from the origin
to the line, $-\infty<\rho<\infty$, and $0 \le \theta \le 2\pi$.

\begin{figure}[ht]
\centering{\epsfig{file=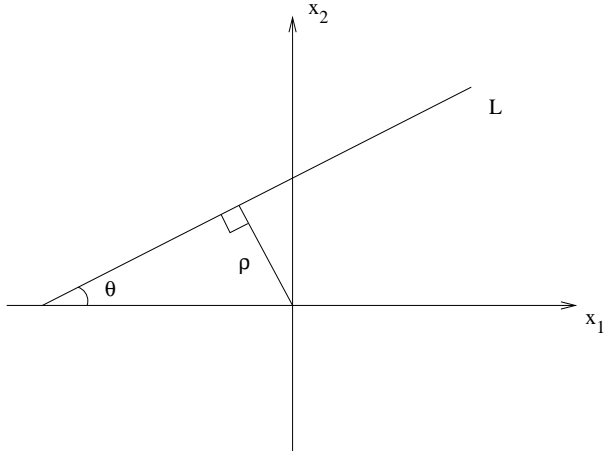,height=6cm,width=8cm}}
\caption{Local coordinates for the mathematical formulation of PET
and SPECT.}
\label{axis}
\end{figure}

\noindent
(ii) The above parameterization implies that, for a fixed $\theta$,
the Cartesian coordinates $(x_1,x_2)$ can be expressed in terms of
the {\it local coordinates} $(\tau,\rho)$ by the equations (see
Section \ref{math})
\begin{equation}
\label{x1x2}
x_1 = \tau\cos\theta - \rho\sin\theta, \quad
x_2 = \tau\sin\theta + \rho\cos\theta.
\end{equation}
A function $f(x_1,x_2)$ rewritten in local coordinates will be denoted by
$F(\tau,\rho,\theta)$,
\[ F(\tau,\rho,\theta) = f(\tau\cos\theta-\rho\sin\theta,
\tau\sin\theta+\rho\cos\theta). \]
Thus $F(\tau,\rho,\theta)$ and $G(\tau,\rho,\theta)$ will denote the
$x$--ray attenuation coefficient $f(x_1,x_2)$ and the distribution of the
radiopharmaceutical $g(x_1,x_2)$, rewritten in local coordinates.
\newline
\noindent
(iii) The line integral of a function $f$ is called its {\it Radon
transform} and will be denoted by $\hat{f}$. In order to compute $\hat{f}$,
we first write $f$ in local coordinates and then integrate with respect
to $\tau$,
\begin{equation}
\label{radon}
\hat f(\rho,\theta) = \int_{-\infty}^{\infty} F(\tau,\rho,\theta)
\mathrm{d} \tau.
\end{equation}
The line integral of the function $g$ with respect to the weight $f$
appearing in equation (\ref{lineint}) is called the {\it attenuated
Radon transform} of $g$ (with the attenuation specified by $f$) and
will be denoted by $\hat g_f$. In order to compute $\hat g_f$, we write
both $g$ and $f$ in local coordinates and then evaluate the following
integral
\begin{equation}
\label{art}
\hat g_f(\rho,\theta) = \int_{-\infty}^{\infty} \mathrm{e}^{- \int^\infty_\tau
F(s,\rho,\theta) \mathrm{d}s} G(\tau,\rho,\theta) \mathrm{d}\tau.
\end{equation}

\subsubsection*{Mathematical Methods}

The basic mathematical problem of both computed tomography and PET
is to reconstruct a function $f$ from the knowledge of
its Radon transform $\hat f$, i.e.\ to solve equation
(\ref{radon}) for $f(x_1,x_2)$ in terms of $\hat f(\rho,\theta)$.
The relevant formula is called the {\it inverse Radon transform}
and is given by
\begin{equation}
\label{irt}
f(x_1,x_2) = \frac{1}{4 \mathrm{i} \pi^2}
(\partial_{x_1}-\mathrm{i}
\partial_{x_2}) \int_{0}^{2 \pi} \mathrm{e}^{\mathrm{i} \theta} \left(
\oint_{-\infty}^{\infty} \frac{\hat{f}(\rho,\theta) \mathrm{d}
\rho}{\rho-(x_2 \cos \theta - x_1 \sin \theta)} \right)
\mathrm{d}\theta,
\end{equation}
where $-\infty<x_j<\infty$, $j=1,2$ and $\oint$ denotes principal
value integral.

A novel approach for deriving equation (\ref{irt}) was introduced in
\cite{foknov}, and is based on the analysis of the equation
\begin{equation}
\label{fnovel}
\left( \frac{1}{2} \left( \lambda+\frac{1}{\lambda} \right)
\partial_{x_1}+ \frac{1}{2\mathrm{i}} \left( \lambda-\frac{1}{\lambda}
\right) \partial_{x_2}\right) \mu(x_1,x_2,\lambda) = f(x_1,x_2),
\end{equation}
where $\lambda$ is a complex parameter different than zero. The
application of this approach to a slight generalization of
equation (\ref{fnovel}) can be used to reconstruct a function
$g$ from the knowledge of its attenuated Radon transform
$\hat g_f$, i.e.\ this approach can be used to solve equation
(\ref{art}) for $g(x_1,x_2)$ in terms of $\hat g_f(\rho,\theta)$
and $f(x_1,x_2)$. The relevant formula, called the {\it inverse
attenuated Radon transform}, was obtained by R.\ Novikov \cite{novi}
by analysing, instead of equation (\ref{fnovel}), the equation
\begin{equation}
\label{gnovel}
\left( \frac{1}{2} \left( \lambda+\frac{1}{\lambda} \right)
\partial_{x_1}+ \frac{1}{2\mathrm{i}} \left( \lambda-\frac{1}{\lambda}
\right) \partial_{x_2} +f(x_1,x_2) \right) \mu(x_1,x_2,\lambda) =
g(x_1,x_2).
\end{equation}

\subsubsection*{Organization of the Paper}

In Section \ref{math} we first review the analysis of equation
(\ref{fnovel}), and then show that if one uses the basic result
obtained in this analysis, it is possible to construct immediately
the inverse attenuated Radon transform. In Section \ref{nume} we
present a new numerical reconstruction algorithm for both PET
and SPECT. This algorithm is based on approximating the given
data in terms of cubic splines. We recall that both the exact
inverse Radon transform as well as the exact inverse attenuated
Radon transform involve the Hilbert transform of the data functions.
For example, the inverse Radon transform involves the function
\begin{equation}
\label{hilbert}
h(\rho,\theta) = \oint_{-\infty}^{\infty}
\frac{\hat{f}(\rho',\theta)}{\rho'-\rho} \mathrm{d} \rho'.
\end{equation}
Existing numerical approaches use the convolution property of the
Fourier transform to compute the Hilbert transform and employ
appropriate filters to eliminate high frequencies. It appears that
our approach has the advantage of simplifying considerably the
mathematical formulas associated with these techniques.
Furthermore, accurate reconstruction is achieved, for noiseless
data, with the additional use of an averaging or of a median filter.
Several numerical tests are presented in Section \ref{tests}. One
of these tests involves the Shepp--Logan phantom \cite{shlo}, see
Figure \ref{petphan}(c).

Numerical algorithms based on the filtered back projection are
discussed in \cite{nattp}--\cite{guno}, while algorithms based on
iterative techniques can be found in \cite{hebert}--\cite{nuyts}.

\section{Mathematical Methods}
\setcounter{equation}{0}
\label{math}

We first review the basic result of \cite{foknov}. It will be
shown later that using this result it is possible to derive both
the inverse Radon as well as the inverse attenuated Radon
transforms in a straightforward manner.

\begin{prop}
Define the complex variable $z$ by
\begin{equation}
\label{defz}
z = \frac{1}{2\mathrm{i}} \left( \lambda-\frac{1}{\lambda} \right) x_1-
\frac{1}{2} \left( \lambda+\frac{1}{\lambda} \right) x_2,
\end{equation}
where $x_1$, $x_2$ are the real Cartesian coordinates $-\infty<x_j<\infty$,
$j=1,2$, and $\lambda$ is a complex variable, $\lambda \ne 0$. Assume that
the function $f(x_1,x_2)$ has sufficient decay as $|x_1|+|x_2| \rightarrow
\infty$. Let $\mu(x_1,x_2,\lambda)$ satisfy the equation
\begin{equation}
\label{mueq}
\frac{1}{2\mathrm{i}} \left( \frac{1}{|\lambda|^2}-|\lambda|^2 \right)
\frac{\partial \mu(x_1,x_2,\lambda)}{\partial \bar{z}} = f(x_1,x_2),
\quad |\lambda| \ne 1, \quad (x_1,x_2) \in \mathbb{R}^2,
\end{equation}
as well as the boundary condition $\mu=\mathrm{O}(1/z)$ as
$|x_1|+|x_2| \rightarrow \infty$. Let $\lambda^+$ and $\lambda^-$
denote the limits of $\lambda$ as it approaches the unit circle
from inside and outside the unit disc respectively, i.e.
\[ \lambda^{\pm} = \lim_{\varepsilon \rightarrow 0} (1 \mp \varepsilon)
\mathrm{e}^{\mathrm{i}\theta}, \quad \varepsilon>0, \quad 0 \le \theta
\le 2\pi. \]

\begin{figure}[ht]
\centering{\epsfig{file=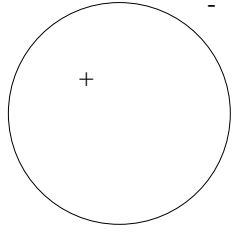,height=3cm,width=3cm}}
\caption{The unit circle.}
\label{circle}
\end{figure}

\noindent
Then
\begin{equation}
\label{finmu}
\mu(x_1,x_2,\lambda^{\pm}) = \mp P^\mp {\hat f}(\rho,\theta) -
\int_\tau^\infty F(\tau',\rho,\theta) \mathrm{d}\tau',
\end{equation}
where $\hat f$ denotes the Radon transform of $f$, $F$ denotes $f$ in
the local coordinates (see the Notation in Section \ref{intro}),
$P^\pm$ denote the usual projection operators in the variable
$\rho$, i.e.
\begin{equation}
\label{p+-}
(P^\pm g)(\rho) = \lim_{\begin{subarray}{c} \varepsilon
\rightarrow 0 \\ \varepsilon >0 \end{subarray}}
\frac{1}{2 \pi \mathrm{i}} \int_{-\infty}^\infty
\frac{g(\rho') \mathrm{d}\rho'}{\rho'-(\rho \pm \mathrm{i}
\varepsilon)} = \pm \frac{g(\rho)}{2} + \frac{1}{2 \pi \mathrm{i}}
\oint_{-\infty}^\infty \frac{g(\rho') \mathrm{d}\rho'}{\rho'-\rho},
\end{equation}
and $\oint$ denotes the principal value integral.
\label{firstprop}
\end{prop}

\noindent
{\bf Proof}.
Before deriving this result, we first note that equation (\ref{defz})
is a direct consequence of equation (\ref{fnovel}). Indeed, equation
(\ref{fnovel}) motivates the introduction of the variable $z$ defined
by equation (\ref{defz}). Taking the complex conjugate of equation
(\ref{defz}) we find
\begin{equation}
\label{conz}
\bar{z} = -\frac{1}{2\mathrm{i}} \left( \bar{\lambda}-\frac{1}{{\bar{\lambda}}}
\right) x_1- \frac{1}{2} \left( \bar{\lambda}+\frac{1}{{\bar{\lambda}}}
\right) x_2.
\end{equation}
Equations (\ref{defz}) and (\ref{conz}) define a change of variables
from $(x_1,x_2)$ to $(z,\bar{z})$. Using this change of variables to
compute $\partial_{x_1}$ and $\partial_{x_2}$ in terms of $\partial_z$
and $\partial_{\bar{z}}$, equation (\ref{fnovel}) becomes (\ref{mueq}).

We now derive equation (\ref{finmu}). The derivation is based on
the following two steps, which have been used extensively in the
field of nonlinear integrable PDEs, see for example \cite{foge}.

(i) In the first step (sometimes called the direct problem), we
consider equation (\ref{mueq}) as an equation which defines $\mu$
in terms of $f$, and we construct an integral representation of $\mu$
in terms of $f$, for {\it all complex} values of $\lambda$. This
representation is
\begin{equation}
\label{repre}
\mu(x_1,x_2,\lambda) = \frac{1}{2 \pi \mathrm{i}} \mathrm{sgn} \!\!\left(
\frac{1}{|\lambda|^2} - |\lambda|^2 \right)
\iint\limits_{\mathbb{R}^2}
\frac{f(x_1',x_2')}{z'-z} \mathrm{d}x_1' \mathrm{d}x_2',
\quad |\lambda| \ne 1.
\end{equation}

Indeed, suppose that the function $\mu(z_R,z_I)$ satisfies the
equation
\[ \frac{\partial \mu(z_R,z_I)}{\partial \bar{z}} = g(z_R,z_I),
\quad z=z_R+\mathrm{i}z_I, \quad -\infty<z_R<\infty,
\quad -\infty<z_I<\infty, \]
as well as the boundary condition $\mu=\mathrm{O}(1/z)$ as $z
\rightarrow \infty$. Then Pompieu's formula (see for example \cite{abfo})
implies
\begin{equation}
\label{pomp}
\mu = -\frac{1}{\pi} \iint\limits_{\mathbb{R}^2} \frac{g(z_R',z_I')}
{z'-z} \mathrm{d}z_R' \mathrm{d}z_I'.
\end{equation}
In our case
\[ g = \frac{2\mathrm{i}f}{\frac{1}{|\lambda|^2} - |\lambda|^2}, \quad
\mathrm{d}z_R \mathrm{d}z_I = \frac{1}{2\mathrm{i}} \left( \frac{1}{|\lambda|^2}
- |\lambda|^2 \right) \mathrm{d}x_1 \mathrm{d}x_2, \]
thus equation (\ref{pomp}) becomes (\ref{repre}).

(ii) In the second step (sometimes called the inverse problem), we
analyze the analyticity properties of $\mu$ with respect to
$\lambda$, and we find an {\it alternative} representation for
$\mu$. This representation involves certain integrals of $f$
called spectral functions. For our problem, this representation is
equation (\ref{finmu}). Indeed, since $\mu$ is an analytic
function of $\lambda$ for $|\lambda| \ne 1$ and since $\mu=
\mathrm{O}(1/\lambda)$ as $\lambda \rightarrow \infty$, we can
reconstruct the function $\mu$ if we know its ``jump'' across the
unit circle:
\begin{equation}
\label{murad}
\mu(x_1,x_2,\lambda) = \frac{1}{2 \pi} \int_0^{2\pi}
\frac{J(x_1,x_2,\theta') \mathrm{e}^{\mathrm{i}
\theta'}}{\mathrm{e}^{\mathrm{i}\theta'}-\lambda}
\mathrm{d}\theta',
\end{equation}
where
\[ J(x_1,x_2,\theta) = \mu(x_1,x_2,\lambda^+)-\mu(x_1,x_2,\lambda^-). \]
Thus we need to compute the limits of $\mu$ as $\lambda$ tends to
$\lambda^\pm$. As $\varepsilon \rightarrow 0$,
\[ \lambda^+ \mp \frac{1}{\lambda^+} \sim (1-\varepsilon)
\mathrm{e}^{\mathrm{i}\theta}
\mp (1+\varepsilon) \mathrm{e}^{-\mathrm{i}\theta}. \]
Substituting this expression in the definition of $z$ (equation
(\ref{defz})) and simplifying, we find
\begin{equation}
\label{zz'}
z'-z \sim (x_1'-x_1) \sin\theta - (x_2'-x_2) \cos\theta +
\mathrm{i} \varepsilon ((x_1'-x_1) \cos\theta + (x_2'-x_2) \sin\theta).
\end{equation}
The right--hand side of this equation can be rewritten in terms of
the local coordinates $\rho$, $\rho'$, $\tau$, $\tau'$: Let
$\mathbf{k}$ and $\mathbf{k}^\perp$ denote two unit vectors along
the line $L$ and perpendicular to this line, respectively. Then
\[ \mathbf{x} = \tau \mathbf{k} + \rho \mathbf{k}^\perp, \]
or
\[ (x_1,x_2) = \tau (\cos\theta,\sin\theta) + \rho (-\sin\theta,
\cos\theta). \]
Hence $x_1$ and $x_2$ are given by equations (\ref{x1x2}). Inverting
these equations we find
\begin{equation}
\label{taurho}
\tau = x_2 \sin\theta + x_1 \cos\theta, \quad
\rho = x_2 \cos\theta - x_1 \sin\theta.
\end{equation}
Thus equation (\ref{zz'}) becomes
\[ z'-z \sim -\rho'+\rho+\mathrm{i} \varepsilon (\tau'-\tau). \]
Substituting this expression in equation (\ref{repre}) and using
the fact that the relevant sign equals 1, we find
\begin{equation}
\label{musim}
\mu(x_1,x_2,\lambda^+) \sim -\frac{1}{2 \pi \mathrm{i}}
\iint\limits_{\mathbb{R}^2}
\frac{f(x_1',x_2') \mathrm{d}x_1' \mathrm{d}x_2'}{\rho'-\rho-\mathrm{i}
\varepsilon (\tau'-\tau)}, \quad \varepsilon \rightarrow 0, \quad
\varepsilon>0.
\end{equation}
Using the change of variables $(x_1,x_2) \leftrightarrow (\tau,\rho)$
defined by equations (\ref{x1x2}) and (\ref{taurho}), and noting that
the relevant Jacobian is 1, i.e.
\[ f(x_1',x_2') \mathrm{d}x_1' \mathrm{d}x_2' = F(\tau',\rho',\theta)
\mathrm{d}\tau' \mathrm{d}\rho', \]
we find that the right--hand side of equation (\ref{musim}) equals
\begin{equation}
\label{intint}
-\frac{1}{2\mathrm{i} \pi}
\iint\limits_{\mathbb{R}^2} \frac{F \mathrm{d}\tau' \mathrm{d}\rho'}
{\rho' - (\rho+\mathrm{i} \varepsilon (\tau'-\tau))}.
\end{equation}

In order to simplify this expression we split the integral over
$\mathrm{d}\tau'$ in the form
\[ \int_{-\infty}^\infty \mathrm{d}\tau' = \int_{-\infty}^\tau
\mathrm{d}\tau' + \int_\tau^\infty \mathrm{d}\tau', \]
and note that in the first integral $\tau'-\tau<0$, while in the second
integral $\tau'-\tau>0$. Thus, using the second set of equations
(\ref{p+-}) the expression in (\ref{intint}) becomes
\[ -\frac{1}{2 \pi \mathrm{i}} \int_{-\infty}^\infty \left( \oint_{-\infty}^\infty
F(\tau',\rho',\theta) \frac{\mathrm{d}\rho'}{\rho'-\rho} \right)
\mathrm{d}\tau' - \frac{1}{2} \int_\tau^\infty F(\tau',\rho,\theta)
\mathrm{d}\tau' + \frac{1}{2} \int_{-\infty}^\tau F(\tau',\rho,\theta)
\mathrm{d}\tau'. \]
Finally, adding and subtracting the integral $\frac{1}{2} \int_\tau^\infty$
we find
\begin{eqnarray*}
\mu(x_1,x_2,\lambda^+) & = & -\frac{1}{2 \pi \mathrm{i}} \int_{-\infty}^\infty \left(
\oint_{-\infty}^\infty F(\tau',\rho',\theta) \frac{\mathrm{d}\rho'}{\rho'-\rho}
\right) \mathrm{d}\tau' \\
& & +\frac{1}{2} \int_{-\infty}^\infty F(\tau',\rho,\theta) \mathrm{d}\tau'-
\int_\tau^\infty F(\tau',\rho,\theta) \mathrm{d}\tau'.
\end{eqnarray*}
The first two terms in the right--hand side of this equation equal
$-P^-\hat{f}$, hence we find (\ref{finmu})$^+$. The derivation of
equation (\ref{finmu})$^-$ is similar. \hfill {\bf QED}

\vskip 0.3cm

Using equation (\ref{finmu}) it is now straightforward to derive
both the inverse Radon and the inverse attenuated Radon
transforms. In this respect we note that the result of Proposition
\ref{firstprop} can be rewritten in the form
\begin{equation}
\label{parbar}
\lim_{\lambda \rightarrow \lambda^\pm} \left\{
\partial_{\bar{z}}^{-1} \left( \frac{f(x_1,x_2)}{\nu(\lambda)}
\right)  \right\} =
\mp P^\mp \hat{f}(\rho,\theta) - \int_\tau^\infty F(\tau',\rho,\theta)
\mathrm{d}\tau',
\end{equation}
where
\begin{equation}
\label{nu}
\nu(\lambda) = \frac{1}{2\mathrm{i}} \left( \frac{1}{|\lambda|^2} - |\lambda|^2
\right).
\end{equation}

\subsubsection*{The Inverse Radon Transform}

Equations (\ref{finmu}) yield
\begin{equation}
\label{capj}
J(x_1,x_2,\theta) = -\frac{1}{\pi \mathrm{i}} \oint_{-\infty}^\infty
\frac{\hat{f}(\rho',\theta)\mathrm{d}\rho'}{\rho'-(x_2 \cos\theta-
x_1 \sin\theta)}.
\end{equation}
Equation (\ref{murad}) implies
\[ \mu(x_1,x_2,\lambda) = \left( -\frac{1}{2 \pi} \int_0^{2\pi}
J(x_1,x_2,\theta) \mathrm{e}^{\mathrm{i} \theta} \mathrm{d}\theta \right)
\frac{1}{\lambda} + \mathrm{O} \left( \frac{1}{\lambda^2} \right). \]
Substituting this expression in equation (\ref{fnovel}) we find
\begin{equation}
\label{fx1x2}
f(x_1,x_2) = \frac{1}{2} (\partial_{x_1}-\mathrm{i} \partial_{x_2}) \left(
-\frac{1}{2 \pi} \int_{0}^{2 \pi} J(x_1,x_2,\theta) \mathrm{e}^{\mathrm{i}
\theta} \mathrm{d}\theta \right).
\end{equation}
Replacing in this equation $J$ by the right--hand side of equation
(\ref{capj}) we find equation (\ref{irt}).

\subsubsection*{The Attenuated Radon Transform}

Equation (\ref{gnovel}) can be rewritten in the form
\[ \frac{\partial \mu}{\partial \bar{z}}+\frac{f}{\nu} \mu =
\frac{g}{\nu}, \]
where $\nu$ is defined by equation (\ref{nu}). Hence
\[ \frac{\partial}{\partial \bar{z}} \left( \mu \exp \!\!\left[
\partial_{\bar{z}}^{-1} \left( \frac{f}{\nu} \right) \right] \right)
= \frac{g}{\nu} \exp \!\!\left[ \partial_{\bar{z}}^{-1} \left(
\frac{f}{\nu} \right) \right], \]
or
\[ \mu \exp \!\!\left[ \partial_{\bar{z}}^{-1} \left( \frac{f}{\nu}
\right) \right] = \partial_{\bar{z}}^{-1} \left( \frac{g}{\nu}
\exp \!\!\left[ \partial_{\bar{z}}^{-1} \left( \frac{f}{\nu}
\right) \right] \right). \]
Replacing in this equation $\partial_{\bar{z}}^{-1} \left( \frac{f}{\nu}
\right)$ by the right--hand side of equation (\ref{parbar}) we find
\[ \mu(x_1,x_2,\lambda^\pm) \mathrm{e}^{\mp P^\mp \hat{f}(\rho,\theta)}
\mathrm{e}^{-\int_\tau^\infty F(\tau',\rho,\theta) \mathrm{d}\tau'} =
\partial_{\bar{z}}^{-1} \left( \frac{g(x_1,x_2)}{\nu(\lambda)}
\mathrm{e}^{\mp P^\mp \hat{f}(\rho,\theta)}
\mathrm{e}^{-\int_\tau^\infty F(\tau',\rho,\theta) \mathrm{d}\tau'} \right). \]
For the computation of the right--hand side of this equation we use again
equation (\ref{parbar}), where $f$ is replaced by $g$ times the two
exponentials appearing in the above relation. Hence
\begin{eqnarray}
\lefteqn{\mu(x_1,x_2,\lambda^\pm) \mathrm{e}^{\mp P^\mp \hat{f}(\rho,\theta)}
\mathrm{e}^{-\int_\tau^\infty F(\tau',\rho,\theta) \mathrm{d}\tau'}=} \nonumber \\
& & {\ds \mp P^\mp \mathrm{e}^{\mp P^\mp \hat{f}(\rho,\theta)} \hat{g}_f(\rho,\theta) -
\int_\tau^\infty G(\tau'\rho,\theta) \mathrm{e}^{\mp P^\mp \hat{f}(\rho,\theta)}
\mathrm{e}^{-\int_{\tau'}^\infty F(s,\rho,\theta) \mathrm{d}s}
\mathrm{d}\tau'.}
\label{p+-new}
\end{eqnarray}
Note that the term $\exp[\mp P^\mp \hat{f}]$ is independent of
$\tau'$, thus this term comes out of the integral
$\int_\tau^\infty$, and furthermore the same term appears in the
left--hand side of equation (\ref{p+-new}). Hence when computing
the jump $\mu(x_1,x_2,\lambda^+)-\mu(x_1,x_2,\lambda^-)$, the
second term in the right--hand side of equation (\ref{p+-new})
cancels and we find that the relevant jump in now given by
\begin{equation}
\label{jump}
J(x_1,x_2,\theta) = -\mathrm{e}^{\int_\tau^\infty F(\tau',\rho,\theta)
\mathrm{d}\tau'} \left( \mathrm{e}^{P^- \hat{f}(\rho,\theta)} P^-
\mathrm{e}^{-P^- \hat{f}(\rho,\theta)} + \mathrm{e}^{-P^+ \hat{f}(\rho,\theta)}
P^+ \mathrm{e}^{P^+ \hat{f}(\rho,\theta)} \right) \hat{g}_f(\rho,\theta)
\end{equation}
where $\tau$ and $\rho$ are expressed in terms of $x_1$ and $x_2$ by
equations (\ref{taurho}).

Equation (\ref{murad}) is still valid, furthermore equation
(\ref{fx1x2}) is valid if $f$ is replaced by $g$. Hence
replacing in equation (\ref{fx1x2}) $f$ by $g$ we find
\begin{equation}
\label{gx1x2}
g(x_1,x_2) = -\frac{1}{4 \pi} (\partial_{x_1}-\mathrm{i}
\partial_{x_2}) \int_{0}^{2 \pi} J(x_1,x_2,\theta)
\mathrm{e}^{\mathrm{i} \theta} \mathrm{d}\theta,
\end{equation}
where $J$ is defined by equation (\ref{jump}). This formula is
equivalent to Novikov's formula.

{\it In summary, let $\hat{g}_f(\rho,\theta)$ be defined by equation
(\ref{art}), let $F(\tau,\rho,\theta)$ denote the function $f(x_1,x_2)$
written in local coordinates (see the Notation) and let $\hat{f}(\rho,
\theta)$ denote the Radon transform of $f(x_1,x_2)$ (see equation
(\ref{radon})). Then $g(x_1,x_2)$ is given by equation (\ref{gx1x2})
where the function $J$ is explicitly given in terms of $\hat{g}_f$ and
$\hat{f}$ by equation (\ref{jump}).}

\section{Reconstruction Algorithm}
\setcounter{equation}{0}
\label{nume}

\subsection{PET Algorithm}
\label{pet}

Taking the real part of equation (\ref{irt}) it follows that $f(x_1,x_2)$
is given by
\begin{equation}
\label{finalf}
f(x_1,x_2) = -\frac{1}{4 \pi^2} \int_0^{2 \pi} h_\rho(\rho,\theta)
\mathrm{d} \theta,
\end{equation}
where $h(\rho,\theta)$ is defined by equation (\ref{hilbert}).

We assume that $f(x_1,x_2)$ has compact support, namely
$f(x_1,x_2)=0$, for ${x_1}^2+{x_2}^2 \ge 1$. For the numerical
calculation of the integral in (\ref{finalf}) we use the formula
\begin{equation}
\label{closed}
\int_{0}^{2 \pi} g(\theta) \mathrm{d}\theta = \frac{2 \pi}{N}
\sum_{i=0}^{N-1} g \left( \frac{2 \pi i}{N} \right).
\end{equation}
Since $g$ is analytic and periodic, this equispaced quadrature
converges at spectral speed \cite{fornberg}. In other words,
(\ref{closed}) represents the optimal quadrature formula for
the above integral and its implementation is likely to result
in high precision even for relatively small values of $N$.
For the numerical calculation of $h_\rho(\rho,\theta)$ we suppose that
$\hat{f}(\rho,\theta)$ is given, for every $\theta$, at $n$
equally spaced points $\rho_i \in [-1,1]$, i.e.\ we suppose that
$\hat{f}_i = \hat{f}(\rho_i,\theta)$ are known. Moreover, in each interval
$[\rho_i,\rho_{i+1}]$ we approximate $\hat{f}(\rho,\theta)$ using
the relation
\begin{equation}
\label{Sappr}
\hat{f}(\rho,\theta)=S_i(\rho,\theta)=A_i \hat{f}_i + B_i
\hat{f}_{i+1} + C_i \hat{f}_i'' + D_i \hat{f}_{i+1}'',
\end{equation}
where
\[ A_i=\frac{\rho_{i+1}-\rho}{\rho_{i+1}-\rho_i}, \,\,\,\, B_i=1-A_i,
\,\,\,\, C_i=\frac{1}{6} ({A_i}^3-A_i) (\rho_{i+1}-\rho_i)^2,
\,\,\,\, D_i=\frac{1}{6} ({B_i}^3-B_i) (\rho_{i+1}-\rho_i)^2, \]
and $\hat{f}_i''$ denotes the second derivative of $\hat{f}(\rho,\theta)$
with respect to $\rho$, at $\rho=\rho_i$. In other words, we approximate
$\hat{f}(\rho,\theta)$ by a cubic spline (in $\rho$) with equally--spaced
nodes. Integrating the spline, we derive a well--known quadrature
formula which, in our setting, reads
\[ h(\rho,\theta) = \sum_{i=1}^{n-1} \int_{\rho_i}^{\rho_{i+1}}
\frac{S_i(\rho',\theta)}{\rho'-\rho} \mathrm{d} \rho'. \]
Following straightforward calculations we obtain
\begin{eqnarray}
h_\rho(\rho,\theta) & = & \sum_{i=1}^{n-1} \left\{
\frac{\hat{f}_{i}}{\rho_{i}-\rho}-\frac{\hat{f}_{i+1}}{\rho_{i+1}-\rho} -
\frac{1}{4} (\rho_i-3 \rho_{i+1}+2 \rho) \hat{f}_i''-
\frac{1}{4} (3 \rho_i-\rho_{i+1}-2 \rho) \hat{f}_{i+1}'' \right. \nonumber \\
& + & \left[ \frac{\hat{f}_i-\hat{f}_{i+1}}{\rho_i-\rho_{i+1}}-
\frac{1}{6} \left( \rho_i-\rho_{i+1}-\frac{3 (\rho_{i+1}-\rho)^2}{\rho_i-\rho_{i+1}}
\right) \hat{f}_i'' \right. \nonumber \\
& + & \left. \left. \frac{1}{6} \left( \rho_i-\rho_{i+1}-\frac{3
(\rho_i-\rho)^2}{\rho_i-\rho_{i+1}} \right) \hat{f}_{i+1}'' \right]
\ln \left| \frac{\rho_{i+1}-\rho}{\rho_i-\rho} \right| \right\}.
\label{newhp}
\end{eqnarray}

In order to calculate numerically $f(x_1,x_2)$ from the data
$\hat{f}(\rho,\theta)$ we first compute the second derivatives
$\hat{f}_i''$. For this purpose we use the subroutine
\texttt{spline} from Numerical Recipes \cite{recipes}, setting
$\hat{f}_1''=\hat{f}_n''=0$ (i.e.\ we use the natural cubic spline
interpolation). Then, for any $x_1$ and $x_2$, we calculate (for
any $\theta$) $\rho$ using (\ref{taurho}b) and
$h_\rho(\rho,\theta)$ using (\ref{newhp}). Finally we calculate
$f(x_1,x_2)$ using (\ref{finalf}).

We note that (\ref{newhp}) contains the term
\[ \ln \left| \frac{\rho_{i+1}-\rho}{\rho_i-\rho} \right|. \]
However, since for the reconstruction the number of the points for $x_1$
and $x_2$ can be different than the number of the $\rho$ points, in general
$\rho \ne \rho_{i+1}$ and $\rho \ne \rho_i$.

\subsection{SPECT Algorithm}
\label{spect}

We denote the first exponential term of the right--hand side of
(\ref{jump}) by $I(\tau,\rho,\theta)$, i.e.
\begin{equation}
\label{int}
I(\tau,\rho,\theta) = \exp\! \left[ \int_\tau^{\sqrt{1-\rho^2}}
F(\tau',\rho,\theta) \mathrm{d}\tau' \right].
\end{equation}
Note that, since we have assumed compact support, the integration
domain is finite, i.e.\ $[\tau,\sqrt{1-\rho^2}]$, and
$F(\tau,\rho,\theta)=0$ for $|\rho| \ge 1$, or for $|\tau| \ge
\sqrt{1-\rho^2}$.

The definitions (\ref{p+-}) become
\[ P^\pm \hat{f}(\rho,\theta) = \pm \frac{1}{2} \hat{f}(\rho,\theta)
- \frac{\mathrm{i}}{2 \pi} h(\rho,\theta). \]
Moreover
\begin{eqnarray*}
& & \exp\! \left[ P^\pm \hat{f}(\rho,\theta) \right] = \exp\! \left[ \pm
\frac{1}{2} \hat{f}(\rho,\theta) \right] \left( \cos
\frac{h(\rho,\theta)}{2 \pi} - \mathrm{i} \sin \frac{h(\rho,\theta)}{2 \pi}
\right), \\
& & \exp\! \left[ -P^\pm \hat{f}(\rho,\theta) \right] = \exp\! \left[ \mp
\frac{1}{2} \hat{f}(\rho,\theta) \right] \left( \cos
\frac{h(\rho,\theta)}{2 \pi} + \mathrm{i} \sin \frac{h(\rho,\theta)}{2 \pi}
\right).
\end{eqnarray*}
We introduce the following notation:
\begin{eqnarray}
& & f^{cpe}(\rho,\theta) = \mathrm{e}^{\frac{1}{2} \hat{f}(\rho,\theta)} \cos
\frac{h(\rho,\theta)}{2 \pi}, \quad
f^{spe}(\rho,\theta) = \mathrm{e}^{\frac{1}{2} \hat{f}(\rho,\theta)} \sin
\frac{h(\rho,\theta)}{2 \pi}, \label{cspe} \\
& & f^{cme}(\rho,\theta) = \mathrm{e}^{-\frac{1}{2} \hat{f}(\rho,\theta)} \cos
\frac{h(\rho,\theta)}{2 \pi}, \quad
f^{sme}(\rho,\theta) = \mathrm{e}^{-\frac{1}{2} \hat{f}(\rho,\theta)} \sin
\frac{h(\rho,\theta)}{2 \pi}, \label{csme} \\
& & f^c(\rho,\theta) = f^{cpe}(\rho,\theta) \hat{g}_f(\rho,\theta), \quad
f^s(\rho,\theta) = f^{spe}(\rho,\theta) \hat{g}_f(\rho,\theta).
\label{cspf}
\end{eqnarray}
Using this notation and setting $R(\tau,\rho,\theta)=
-J(\tau,\rho,\theta)$, after some calculations, equation (\ref{jump})
becomes
\begin{equation}
\label{rel2g}
R(\tau,\rho,\theta) = I(\tau,\rho,\theta) \left(
(f^{cme} - \mathrm{i} f^{sme}) (P^-f^c+\mathrm{i} P^-f^s)+
(f^{cme} + \mathrm{i} f^{sme}) (P^+f^c-\mathrm{i} P^+f^s) \right).
\end{equation}

We now set
\[ \oint_{-\infty}^\infty \frac{f^c(\rho',\theta)}{\rho'-\rho}
\mathrm{d} \rho' = h^c(\rho,\theta), \quad
\oint_{-\infty}^\infty \frac{f^s(\rho',\theta)}{\rho'-\rho}
\mathrm{d} \rho' = h^s(\rho,\theta), \]
thus equation (\ref{rel2g}) becomes
\[ R(\tau,\rho,\theta) = -\mathrm{i} I(\tau,\rho,\theta) \left(
f^{cme} \left( \frac{1}{\pi} h^c + 2 f^s \right) +
f^{sme} \left( \frac{1}{\pi} h^s - 2 f^c \right) \right). \]
We denote the right--hand side of this equation by
$-\mathrm{i} r(\tau,\rho,\theta)$.
Taking the real part of $g(x_1,x_2)$ in (\ref{gx1x2}), we obtain
\begin{equation}
\label{finalp}
g(x_1,x_2) = \frac{1}{4 \pi} \int_0^{2 \pi} \left( r_{x_1} \sin \theta
- r_{x_2} \cos \theta \right) \mathrm{d} \theta,
\end{equation}
where $\tau$ and $\rho$ are given by (\ref{taurho}) and
\begin{equation}
\label{relg}
r(\tau,\rho,\theta) = I(\tau,\rho,\theta) \left(
f^{cme} \left( \frac{1}{\pi} h^c + 2 f^s \right) +
f^{sme} \left( \frac{1}{\pi} h^s - 2 f^c \right) \right).
\end{equation}

For the numerical calculation of the Hilbert transform we write
\begin{eqnarray}
h(\rho,\theta) & = & \int_{-1}^1
\frac{\hat{f}(\rho,\theta)}{\rho'-\rho} \mathrm{d} \rho'+
\int_{-1}^1 \frac{\hat{f}(\rho',\theta)- \hat{f}(\rho,\theta)}
{\rho'-\rho} \mathrm{d} \rho' \nonumber \\
& = & \hat{f}(\rho,\theta) \ln\! \left(
\frac{1-\rho}{1+\rho} \right) + \sum_{i=1}^{n-1} \int_{\rho_i}^{\rho_{i+1}}
\frac{S_i(\rho',\theta)-\hat{f}(\rho,\theta)}{\rho'-\rho} \mathrm{d} \rho'.
\label{newnewh}
\end{eqnarray}
If $\rho=\rho_i$ or $\rho=\rho_{i+1}$ the integral in the right--hand
side of (\ref{newnewh}) can be written
\[ \int_{\rho_i}^{\rho_{i+1}}
\frac{S_i(\rho',\theta)-S_i(\rho,\theta)}{\rho'-\rho} \mathrm{d}
\rho'. \]
Thus, after some calculations, we obtain
\begin{eqnarray}
& & \int_{\rho_i}^{\rho_{i+1}}
\frac{S_i(\rho',\theta)-\hat{f}(\rho,\theta)}{\rho'-\rho} \mathrm{d}
\rho' = -\hat{f}_i+\hat{f}_{i+1} \nonumber \\ & & + \frac{1}{36}
\left( 4 {\rho_i}^2 - 5 \rho_i \rho_{i+1} - 5 {\rho_{i+1}}^2 - 3
(\rho_i-5 \rho_{i+1}) \rho -6 \rho^2 \right) \hat{f}_i'' \nonumber \\
& & + \frac{1}{36} \left( 5 {\rho_i}^2 + 5 \rho_i \rho_{i+1} - 4
{\rho_{i+1}}^2 - 3 (5 \rho_i-\rho_{i+1}) \rho +6 \rho^2 \right)
\hat{f}_{i+1}''.
\label{newint1}
\end{eqnarray}
If $\rho \ne \rho_i$ and $\rho \ne \rho_{i+1}$ the integral in the
right--hand side of (\ref{newnewh}) can be written
\[ \int_{\rho_i}^{\rho_{i+1}} \frac{S_i(\rho',\theta)}{\rho'-\rho}
\mathrm{d} \rho' - \hat{f}(\rho,\theta) \ln \left|
\frac{\rho_{i+1}-\rho}{\rho_i-\rho} \right|, \]
and after some calculation we obtain
\begin{eqnarray}
h(\rho,\theta) & = & \sum_{i=1}^{n-1} \left\{ F_i -
\frac{1}{\rho_i-\rho_{i+1}} \ln \left|
\frac{\rho_{i+1}-\rho}{\rho_i-\rho} \right| \left[
(\rho_{i+1}-\rho) \hat{f}_i - (\rho_{i}-\rho) \hat{f}_{i+1}
\right. \right.
\nonumber \\
& - & \left. \left.  \frac{1}{6} (\rho_{i}-\rho) (\rho_{i+1}-\rho)
\left( (\rho_{i}-2 \rho_{i+1}+\rho) \hat{f}_i'' + (2 \rho_{i} -
\rho_{i+1} - \rho) \hat{f}_{i+1}'' \right) \!\right] \!\right\} \!\!,
\label{newint2}
\end{eqnarray}
where $F_i$ is the right--hand side of (\ref{newint1}).

In order to calculate numerically $I(\tau,\rho,\theta)$ for any
$x_1$, $x_2$, $\theta$, we use relations (\ref{finalf}) and
(\ref{taurho}b). Thus
\[ f(x_1,x_2) = -\frac{1}{4 \pi^2} \int_0^{2 \pi} h_\rho
(x_2 \cos t-x_1 \sin t,t) \mathrm{d}t, \]
and consequently
\begin{equation}
\label{capf}
F(\tau,\rho,\theta) = -\frac{1}{4 \pi^2} \int_0^{2 \pi}
h_\rho (\tau \sin (\theta-t) + \rho \cos (\theta-t),t)
\mathrm{d}t,
\end{equation}
where $\tau$ and $\rho$ are given from (\ref{taurho}) and $h_\rho$
from (\ref{newhp}). We can now calculate $F(\tau,\rho,\theta)$
following the procedure outlined in the previous section. We then
calculate $I(\tau,\rho,\theta)$ using relation (\ref{int}) if
$\tau \ge 0$, alternatively the relation
\begin{equation}
\label{intminus}
I(\tau,\rho,\theta) = \exp\! \left[ \hat{f}(\rho,\theta)-
\int_{-\sqrt{1-\rho^2}}^\tau F(\tau',\rho,\theta) \mathrm{d}\tau'
\right]
\end{equation}
if $\tau<0$. For the numerical calculation of the integrals appearing in
(\ref{int}) and (\ref{intminus}) we use the Gauss--Legendre quadrature
with two functional evaluations at every step, i.e.
\[ \int_\alpha^\beta F(\tau',\rho,\theta) \mathrm{d}\tau'\approx
w_1 F(\tau_1,\rho,\theta)+w_2 F(\tau_2,\rho,\theta), \]
where the abscissas $\tau_1$, $\tau_2$ and the weights $w_1$, $w_2$ are
given by
\[ \tau_1=\alpha+(\beta-\alpha) \left( \frac12 -\frac{\sqrt{3}}{6} \right),
\quad \tau_2=\alpha+(\beta-\alpha) \left(\frac12 +\frac{\sqrt{3}}{6} \right),
\quad w_1=w_2=\frac12(\beta-\alpha). \]
We also notice that we have tried subdivision of the interval
$(\alpha,\beta)$ into several intervals and the improvement is
very minor. Therefore we use just one interval, i.e.\ two function
evaluations per quadrature, since the major increase in running
time of the program implicit in using panel quadrature is not
justified by the modest improvement in accuracy.

For the numerical calculation of the integrals in (\ref{finalp})
and (\ref{capf}) we use again formula (\ref{closed}), resulting in
spectral convergence. For the numerical calculation of the partial
derivatives $r_{x_1}$ and $r_{x_2}$ in (\ref{finalp}) we use the
forward difference scheme
\[ f'(x) \approx \frac{-3f(x)+4f(x+\Delta x)-f(x+2 \Delta x)}{2 \Delta
x} \]
for the first half of the interval $[-1,1]$, and the backward difference
scheme
\[ f'(x) \approx \frac{3f(x)-4f(x-\Delta x)+f(x-2 \Delta x)}{2 \Delta
x} \]
for the second half.

Thus, for the numerical calculation of $g(x_1,x_2)$ from the data
$\hat{f}(\rho,\theta)$ and $\hat{g}_f(\rho,\theta)$ we apply the
following procedure: First we calculate the second derivatives
$\hat{f}_i''$, using subroutine \texttt{spline}. Consequently, we
calculate $h(\rho,\theta)$ using (\ref{newnewh}) and
(\ref{newint1}) for all given $\rho$ and $\theta$. We note that if
$|\rho_i|=1$, then, since we have assumed compact support,
$\hat{f}(\rho,\theta)=0$, thus the first term in (\ref{newnewh})
is absent. We then calculate $f^{cpe}(\rho,\theta)$ and
$f^{spe}(\rho,\theta)$ using (\ref{cspe}), as well as
$f^c(\rho,\theta)$ and $f^s(\rho,\theta)$ using (\ref{cspf}) (at
this stage we use the second data function $\hat{g}_f$). Finally
we calculate, again using \texttt{spline}, the second derivatives
for the natural cubic spline interpolation of the functions
$f^c(\rho,\theta)$ and $f^s(\rho,\theta)$.

Having calculated all the necessary second derivatives we now
proceed as follows: First we calculate $\hat{f}(\rho,\theta)$ for
any $x_1$, $x_2$ (and $\theta$) using (\ref{taurho}) and
(\ref{Sappr}). For this purpose we have used subroutine
\texttt{splint} from Numerical Recipes. Consequently we calculate
$h(\rho,\theta)$ using (\ref{newint2}). Then we calculate
$f^{cme}(\rho,\theta)$ and $f^{sme}(\rho,\theta)$ using
(\ref{csme}), $f^c(\rho,\theta)$ and $f^s(\rho,\theta)$ using
\texttt{splint} and finally $h^c(\rho,\theta)$ and
$h^s(\rho,\theta)$ using relations similar to (\ref{newint2}).
These last six functions are used in (\ref{relg}). We then
calculate $I(\tau,\rho,\theta)$ as described earlier. Finally we
calculate $r(\tau,\rho,\theta)$ using (\ref{relg}) and
consequently $g(x_1,x_2)$ using (\ref{finalp}).

\section{Numerical Tests}
\setcounter{equation}{0}
\label{tests}

The $\theta$ points are equally spaced in $[0,2 \pi]$, while the
$\rho$ points are equally spaced in $[-1,1]$. The density plots
presented below were drawn by using \texttt{Mathematica}
\cite{math}. The dark color represents zero (or negative) values
while the white color represents the maximum value of the original
(or reconstructed) function.

\begin{figure}[ht!]
\centering{\epsfig{file=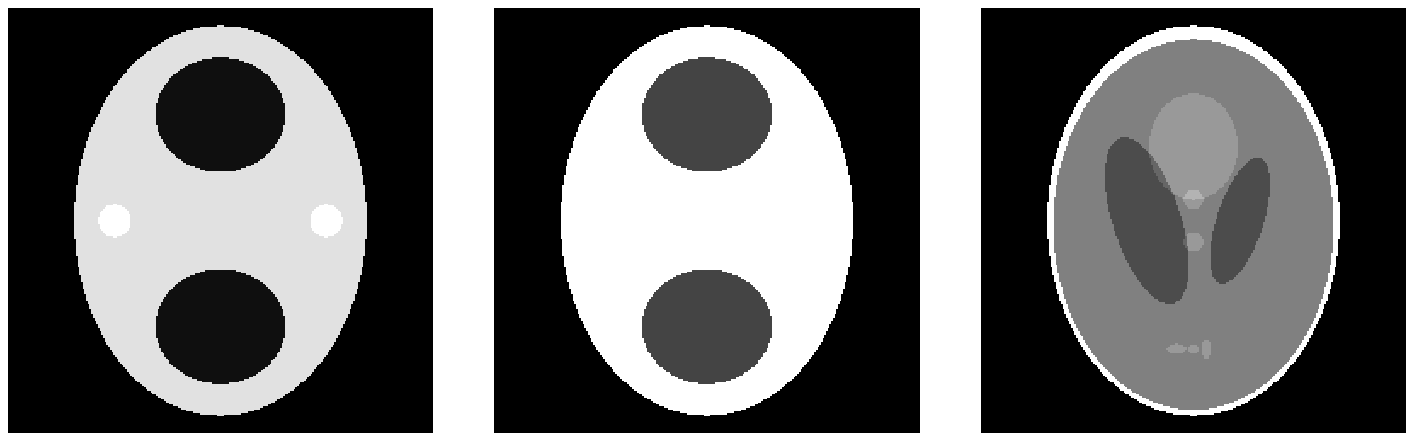}}
\vskip -0.3cm
(a) \hskip 4.3cm (b) \hskip 4.3cm (c)
\caption{Test phantoms for the PET algorithm.}
\label{petphan}

\vskip 0.3cm

\centering{\epsfig{file=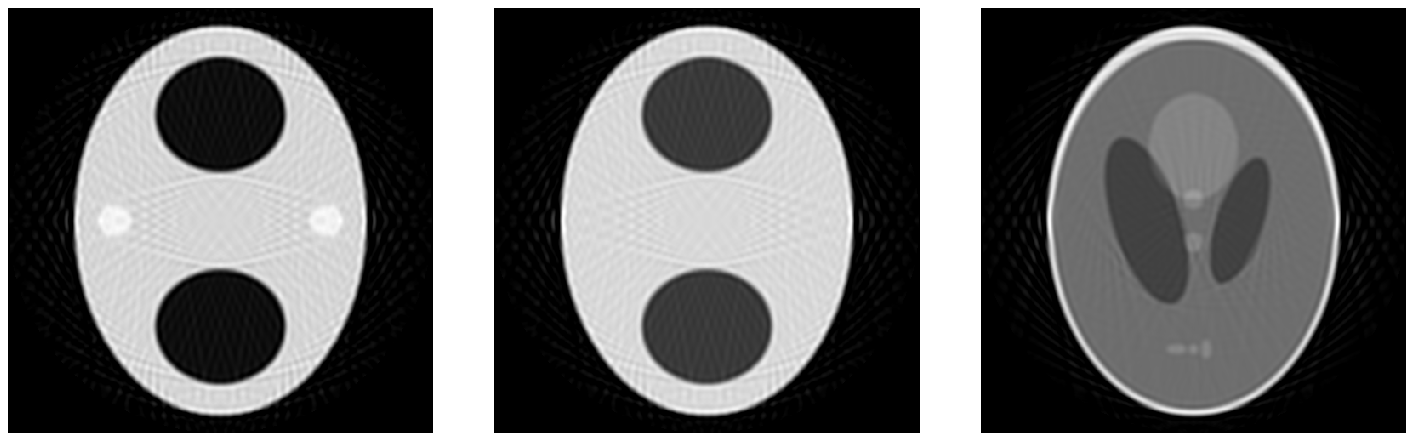}}
\vskip -0.3cm
(a) \hskip 4.3cm (b) \hskip 4.3cm (c)
\caption{The reconstruction of the phantoms of Figures \ref{petphan}
before the filtering procedure.}
\label{petphan1}

\vskip 0.3cm

\centering{\epsfig{file=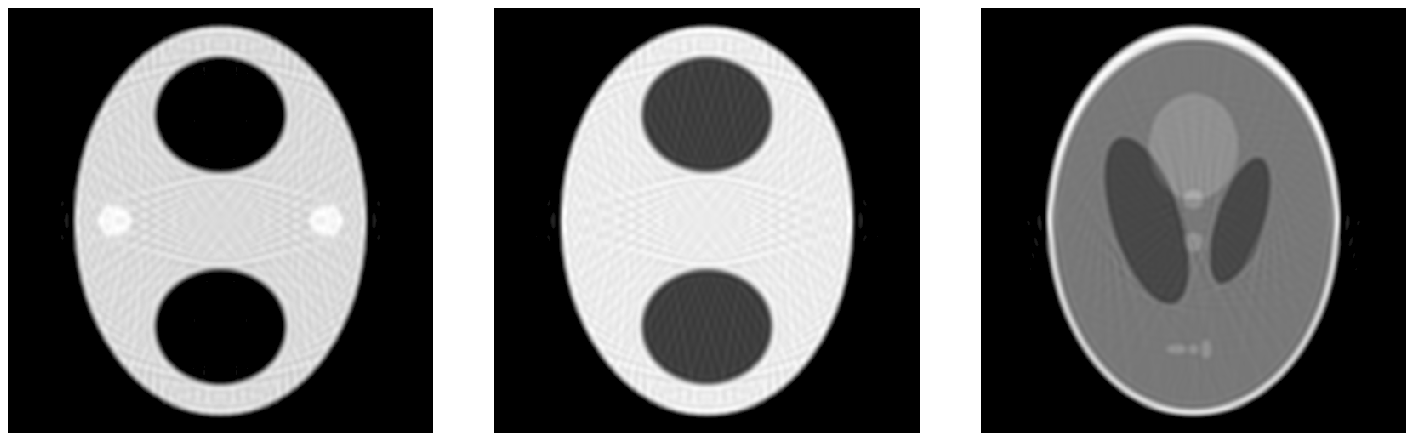}}
\vskip -0.3cm
(a) \hskip 4.3cm (b) \hskip 4.3cm (c)
\caption{The reconstruction of the phantoms of Figures \ref{petphan}
after the filtering procedure.}
\label{petphan2}
\end{figure}

First we tested the PET algorithm for the three different phantoms
shown in Figures \ref{petphan}. Figures (a) and (b) were taken
from \cite{kuny} and \cite{guno}, respectively. These figures
depict the attenuation coefficient for a function $f(x_1,x_2)$
modelling a section of a human thorax. The small circles represent
bones and the larger ellipses the lungs. Figure (c) is the well
known Shepp--Logan phantom, which provides a model of a head
section. All these phantoms consist of different ellipses with
various densities.

Using the Radon transform (\ref{radon}), we computed the data
function $\hat f(\rho,\theta)$ for 200 points for $\theta$ and 100
points for $\rho$. This computation was carried out by using
\texttt{Mathematica}. We then used these data in the numerical
algorithm to reevaluate $f(x_1,x_2)$. Furthermore, in order to
remove the effect of the Gibbs--Wilbraham phenomenon, we applied
an averaging filter as follows: We first found the maximum value
($\max$) of $f(x_1,x_2)$ in the reconstructed image. We then set
to zero those values of $f(x_1,x_2)$ which were less than
$\frac{1}{20}\max$. Finally we applied the averaging filter with
averaging parameter $a=0.005$. This filtering procedure was
applied five times, with the additional elimination of those
values of $f(x_1,x_2)$ which were less than $\frac{1}{20}\max$ at
the end of the procedure. In Figures \ref{petphan1} and
\ref{petphan2} we present the results before and after the
filtering procedure, respectively. The reconstruction took place
in a $500 \times 500$ grid.

\begin{figure}[ht!]
\centering{\epsfig{file=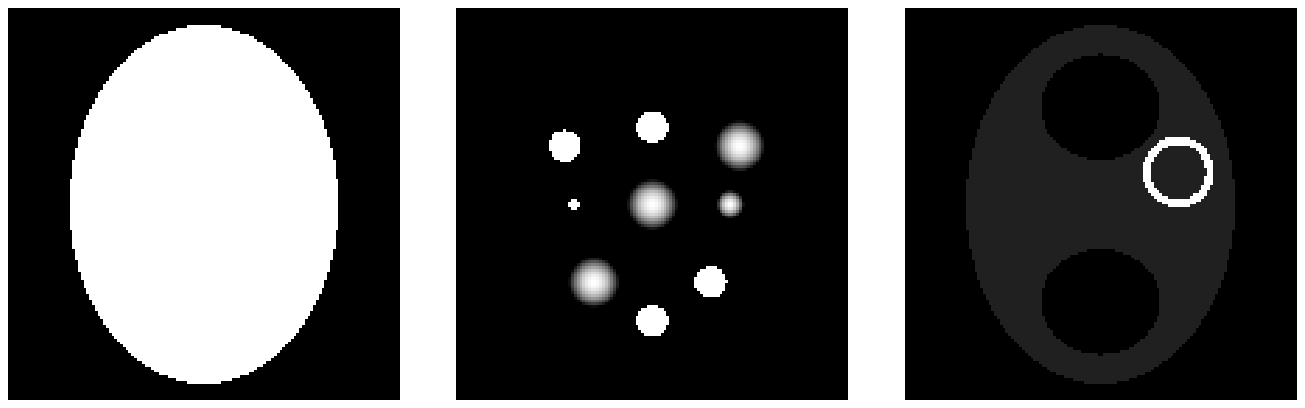}}
\vskip -0.3cm
(a) \hskip 3.9cm (b) \hskip 3.9cm (c)
\caption{Test phantoms for the SPECT algorithm.
In Figures (a) and (b) the function $f(x_1,x_2)$ is given
by Figure \ref{petphan}(a), while in Figure (c) the function
$f(x_1,x_2)$ is given by Figure \ref{petphan}(b).}
\label{spectphan}

\vskip 0.3cm

\centering{\epsfig{file=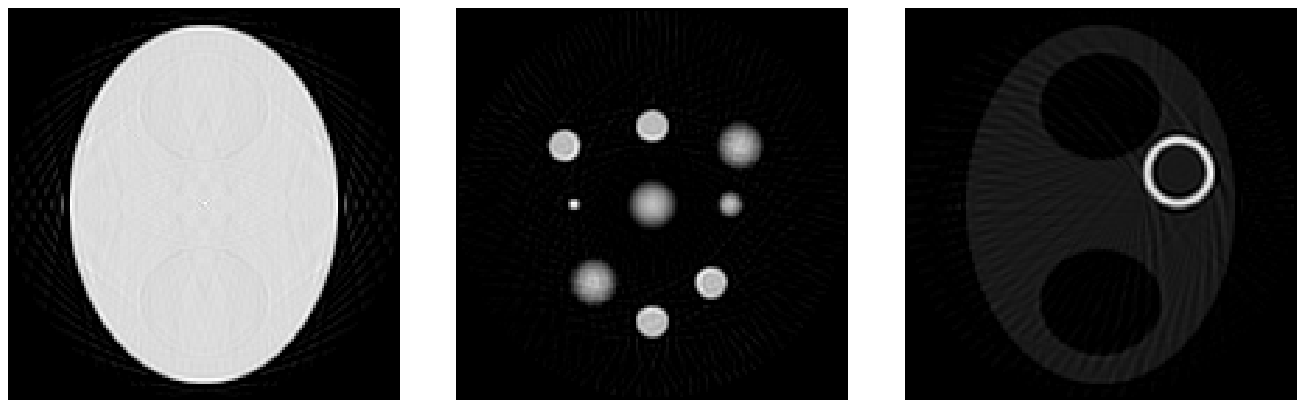}}
\vskip -0.3cm
(a) \hskip 3.9cm (b) \hskip 3.9cm (c)
\caption{The reconstruction of the phantoms of Figures \ref{spectphan}
before the filtering procedure.}
\label{spectphan1}

\vskip 0.3cm

\centering{\epsfig{file=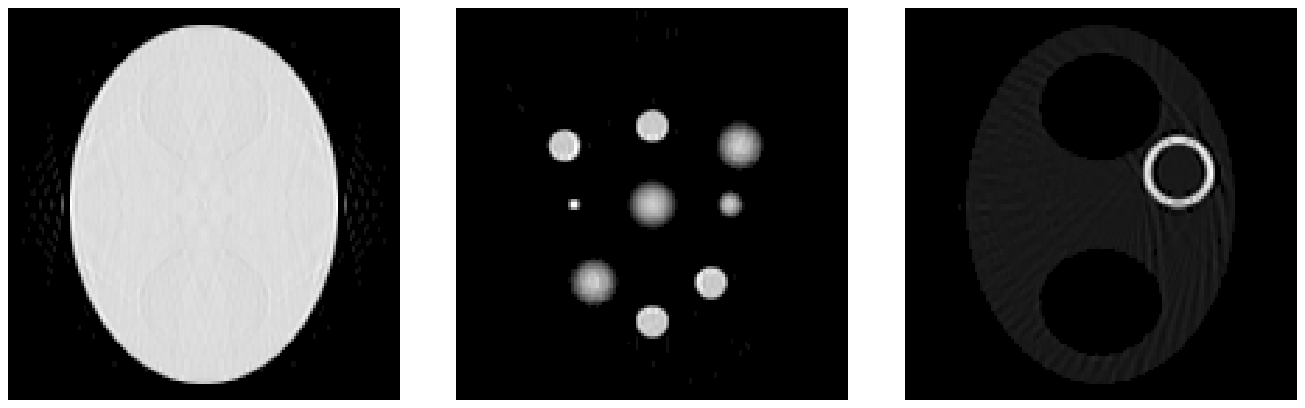}}
\vskip -0.3cm
(a) \hskip 3.9cm (b) \hskip 3.9cm (c)
\caption{The reconstruction of the phantoms of Figures \ref{spectphan}
after the filtering procedure.}
\label{spectphan2}
\end{figure}

We then tested the SPECT algorithm for the three different
phantoms shown in Figures \ref{spectphan}. Figures (a) and (b)
were taken from \cite{kuny}. In these cases the function
$f(x_1,x_2)$ is given by Figure \ref{petphan}(a). Figure (c) was
taken from \cite{guno}. The white ring represents the distribution
of the radiopharmaceutical at the myocardium. In this case the
function $f(x_1,x_2)$ is given by Figure \ref{petphan}(b).

By using the Radon transform (\ref{radon}), and the attenuated
Radon transform (\ref{art}), we computed the data functions $\hat
f(\rho,\theta)$ and $\hat g_f(\rho,\theta)$ for 200 values of
$\theta$ and 100 points of $\rho$ (again using
\texttt{Mathematica}). We consequently used these data in our
program to re--evaluate $g(x_1,x_2)$. In order to remove the
effect of the Gibbs--Wilbraham phenomenon, a median filter was
used, with the additional elimination of those values of
$g(x_1,x_2)$ which were less than $\frac{1}{20}\max$ before and
after the application of the filter. The results are shown in
Figures \ref{spectphan1} and \ref{spectphan2}, before and after
the filtering procedure respectively. The reconstruction took
place in a $140 \times 140$ grid.

For the above phantoms it seems that even a rough estimation of
$F(\tau,\rho,\theta)$ is sufficient for an accurate
reconstruction. This means that, in order to compute numerically
$F(\tau,\rho,\theta)$ using (\ref{capf}), it is sufficient to use
ten equally spaced points for $t$, rather than $200$. This
reduces considerably the reconstruction time.

\section*{Acknowledgments}

\noindent
V.M.\ was supported by a Marie Curie Individual Fellowship of the
European Community under contract number HPMF-CT-2002-01597. We are
grateful to Professor B.\ Hutton for useful suggestions.

\end{document}